\documentclass[11pt]{article}
\usepackage{graphicx}
\usepackage[margin=3cm]{geometry}
\usepackage[authoryear]{natbib}
\usepackage{amssymb,amsmath,amsthm}

\newcommand{\N}{\ensuremath{\mathbb{N}}}
\newcommand{\Z}{\ensuremath{\mathbb{Z}}}
\newcommand{\R}{\ensuremath{\mathbb{R}}}
\newcommand{\PP}{\ensuremath{\mathbb{P}}}
\newcommand{\E}{\ensuremath{\mathbb{E}}}
\newcommand{\I}{\ensuremath{\mathbb{I}}}

\newtheorem{theorem}{Theorem}

\title{A simple method for implementing Monte Carlo tests}
\author{Dong Ding, Axel Gandy, and Georg Hahn
  \\Department of Mathematics, Imperial College London}
\date{}

\begin{document}
\maketitle

\begin{abstract}
We consider a statistical test whose p-value can only be
  approximated using Monte Carlo simulations. We are interested in
  deciding whether the p-value for an observed data set lies above or
  below a given threshold such as 5\%. We want to ensure that the
  resampling risk, the probability of the (Monte Carlo) decision being
  different from the true decision, is uniformly bounded. This article
  introduces a simple open-ended method with this property, the
  confidence sequence method (CSM). We compare our approach to another
  algorithm, SIMCTEST, which also guarantees an (asymptotic) uniform
  bound on the resampling risk, as well as to other Monte Carlo
  procedures without a uniform bound.  CSM is free of tuning parameters and conservative.
  It has the same theoretical
  guarantee as SIMCTEST and, in many settings, similar stopping
  boundaries.  As it is much simpler than other methods, CSM is a
  useful method for practical applications.
\end{abstract}

\section{Introduction}
\label{section_introduction}
Suppose we want to use a one-sided statistical test with null hypothesis $H$
based on a test statistic $T$ with observed value $t$.  We aim to calculate the p-value
$$p = \PP(T \geq t),$$
where the measure $\PP$ is ideally the true null distribution in a simple hypothesis. Otherwise, it can be an estimated distribution in a bootstrap scheme, or a distribution conditional on an ancillary statistic, etc. 

We consider the scenario in which $p$ cannot be evaluated in closed
form, but can be approximated using Monte Carlo simulation, e.g.\
through bootstrapping or drawing permutations.  To be precise, we
assume we can generate a stream $(T_i)_{i \in \N}$ of i.i.d.\ random
variables from the distribution of a test statistic $T$ under $\PP$.
The information about whether or not $T_i$ exceeds the observed value
$t$ is contained in the random variable $X_i = \I(T_i \geq t)$, where
$\I$ denotes the indicator function. It is a Bernoulli random variable
satisfying $\PP(X_i=1)=p$. We will formulate algorithms in terms of
$X_i$.

\cite{gleser1996bootstrap} suggests that two individuals using the same statistical method on
the same data should reach the same conclusion.  For tests, the
standard decision rule is based on comparing $p$ to a threshold
$\alpha$.  In the setting we consider, Monte Carlo methods are used to
compute an estimate $\hat p$ of $p$, which is then compared to
$\alpha$ to reach a decision.
 
 We are interested in procedures which provide a
 user-specified uniform bound $\epsilon>0$ on the probability of the
 decision based on $\hat p$ being different to the decision based on $p$. We
 call this probability the resampling risk and define it formally as
 $$\text{RR}_p(\hat{p}) =
 \begin{cases}
 \PP_p(\hat{p} > \alpha) & \text{if }p \leq \alpha,\\
 \PP_p(\hat{p} \leq \alpha) & \text{if }p > \alpha,
 \end{cases}$$
 where $\hat p$ is the p-value estimate computed by a Monte Carlo procedure. 
 We are looking for procedures that achieve
 \begin{equation}
 \label{eq:unifboundRR}
 \sup_{p \in [0,1]} \text{RR}_p(\hat{p}) \leq \epsilon
 \end{equation}
 for a pre-specified $\epsilon>0$. 
 
 We introduce a simple sequential Monte Carlo testing procedure achieving
 \eqref{eq:unifboundRR} for any $\epsilon>0$, which we call the {\it
 confidence sequence method} (CSM). Our method is based on a
 confidence sequence for $p$, that is a sequence of (random) intervals
 with a joint coverage probability of at least $1-\epsilon$.
 We will use the sequences constructed in \cite{Robbins1970,Lai1976}.
 A decision whether to reject $H$ is reached as soon as the first
 interval produced in the confidence sequence ceases to contain the
 threshold $\alpha$.

The basic (non-sequential) Monte Carlo estimator \citep{davison1997bootstrap}
 $$\hat{p} = \frac{1 +  \sum^n_{i=1} X_i}{1+n}$$
 does not guarantee a small uniform bound on $\text{RR}_p(\hat{p})$, where $n$ is a pre-specified number of Monte Carlo simulations. In
 fact, the lowest uniform bound on the resampling risk for this
 estimator is at least $0.5$ \citep{Gandy2009}.
 
 A variety of procedures for sequential Monte Carlo testing are available in the literature which target different error measures. \cite{Silva2009, SilvaAssuncao2013} bound the power loss of the test while minimising the expected number of steps.
 \citet[Section 4]{silva2018truncated} construct truncated sequential Monte Carlo algorithms which bound the power loss and the level of significance in comparison to the exact test by arbitrarily small numbers. Other algorithms aim to control the resampling risk \citep{FayFollmann2002, Fay2007, Gandy2009, Kim2010}. \cite{Fay2007} use a truncated sequential probability ratio test (SPRT) boundary and discuss the resampling risk, but do not aim for a uniform lower bound on it. \cite{FayFollmann2002} and \cite{Kim2010} ensure a uniform bound on the resampling risk under the assumption that the random variable $p$ belongs to a certain class of distributions. Besides being a much less restrictive requirement than \eqref{eq:unifboundRR}, one drawback of this approach is that in real situations, the distribution of $p$ is typically not fully known, as
 this would require knowledge of the underlying true sampling distribution.
 
 We mainly compare our method to the existing approach of \citet{Gandy2009},
 which we call SIMCTEST in the present article.  SIMCTEST works on the
 partial sum $S_n=\sum_{i=1}^n X_i$ and reaches a decision on $p$ as
 soon as $S_n$ crosses suitably constructed decision boundaries.  SIMCTEST is
 specifically constructed to guarantee a desired uniform bound on the
 resampling risk.
 
 Procedures for Monte Carlo testing can be classified as open-ended
 and truncated procedures \citep{SilvaAssuncao2013}. A truncated
 approach \citep{davidson2000bootstrap, silva2018truncated,
   BesagClifford1991, SilvaAssuncao2013} specifies a maximum number of
 Monte Carlo simulations in advance and forces a decision before or at
 the end of all simulations.  Open-ended procedures
 (e.g., \cite{Gandy2009}) do not have a stopping rule or an upper
 bound on the number of steps.  Open-ended procedures can be turned
 into truncated procedures by forcing a decision after a fixed number
 of steps. Truncated procedures cannot guarantee a uniform bound on
 the resampling risk -- Section~\ref{section_RR_comparison}
 demonstrates this.
 
 In practice, a truly open-ended procedure will often not be feasible.
 This is obvious in settings where generating samples is very
 time-consuming \citep{tango2005flexibly,kulldorff2001prospective}, but
 it is also true in other settings where a large number of samples is being
 generated, as infinite computational effort is never available.
 
 In the related context of Monte Carlo tests for multiple comparisons,
 algorithms guaranteeing a uniform error bound on any erroneous decision
 have  been considered \citep{GandyHahn2014,GandyHahn2016framework}.
 Though not explicitly discussed,
 CSM could in fact be considered a special case of the framework of \citet{GandyHahn2016framework}
 when testing one hypothesis using a Bonferroni correction \citep{Bonferroni1936}.
 
 This article is structured as follows. We first describe the
 confidence sequence method in Section~\ref{sec:CSM} before briefly
 reviewing SIMCTEST in Section~\ref{subsection_method_gandy}.  We
 compare the (implied) stopping boundaries of both methods and
 investigate the real resampling risk incurred from their use in
 Section~\ref{section_stopping_boundaries}.
 In Section~\ref{section_dominating_seq}, we investigate the rate at which both
 methods spend the resampling risk and construct a new spending
 sequence for SIMCTEST which (empirically) gives uniformly tighter
 stopping boundaries than CSM, thus leading to faster decisions on $p$.
 In Section~\ref{section_RR_comparison}, we show that neither procedure bounds the resampling risk in the truncated Monte Carlo setting, but that the truncated version of CSM performs well compared to more complicated algorithms.
 The article concludes with two example applications in Section~\ref{section_example} and a discussion in Section~\ref{section_discussion}.
 
\begin{figure}
\centering
\resizebox*{10cm}{!}{\includegraphics{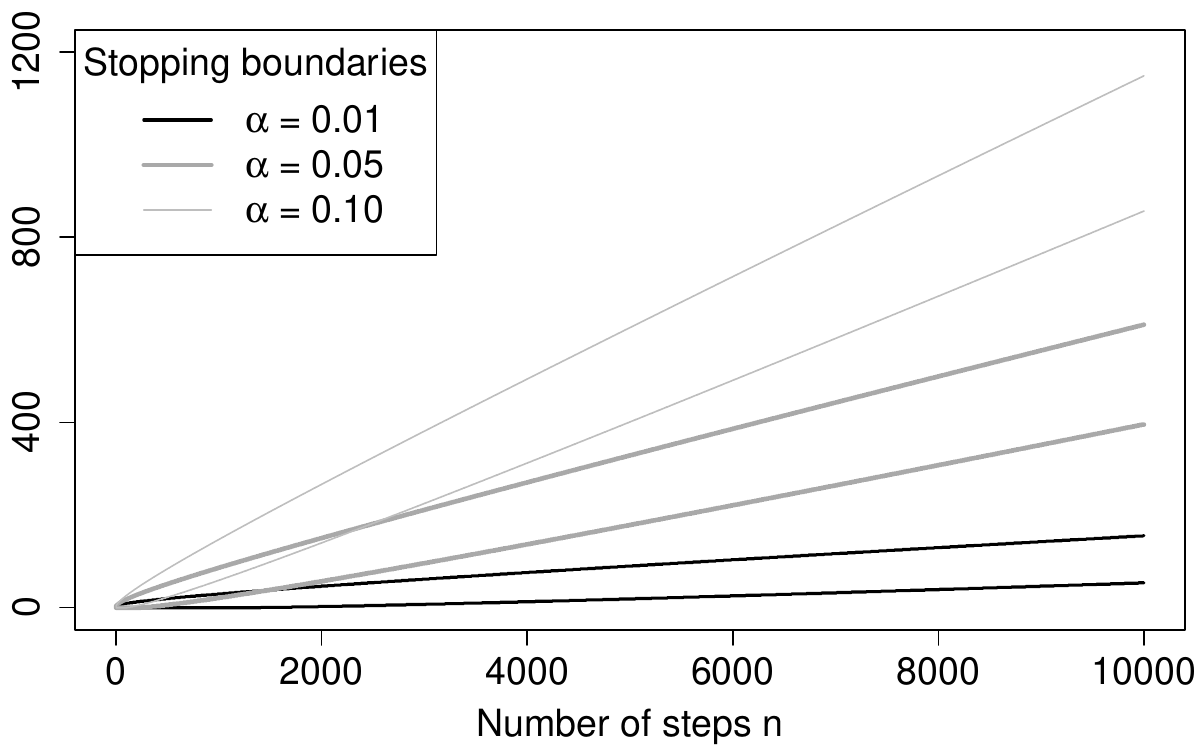}}\hspace{5pt}
\caption{Lower ($l_n$) and upper ($u_n$) stopping boundaries of CSM for several thresholds $\alpha$.} 
\label{fig:dif_alpha_bounds}
\end{figure}

\section{The Confidence Sequence Method}
\label{sec:CSM}
This section introduces the confidence sequence method (CSM),
a simple algorithm to compute a decision on $p$ while bounding the resampling risk.

Let $\epsilon\in (0,1)$ be the desired bound on the resampling risk.
Using independent Bernoulli$(p)$ random variables $X_i$, $i \in \N$,
the following inequality holds \citep{Robbins1970}:
\begin{align}
\PP_p \left( \exists n \in \N: b(n,p,S_n) \leq \frac{\epsilon}{n+1} \right) \leq \epsilon
\label{eqn:lai_original}
\end{align}
for all $p \in (0,1)$, where $b(n,p,x) = \binom{n}{x} p^x (1-p)^{n-x}$
and $S_n = \sum^n_{i=1} X_i$. Then,
$I_n=\{p\in [0,1]:(n+1)b(n,p,S_n) >\epsilon\}$ is a sequence of
confidence sets that has the desired joint coverage probability
$1-\epsilon$.

The $I_n$ are intervals \citep{Lai1976}. Indeed, if $0<S_n<n$
we obtain $I_n=(g_n(S_n),f_n(S_n))$, where $g_n(x) < f_n(x)$ are the two
distinct roots of $(n+1)b(n,p,x) = \epsilon$.  If $S_n=0$ then the
equation $(n+1)b(n,p,0) = \epsilon$ has only one root $r_n$, leading to
$I_n = [0,r_n)$.  Likewise for $S_n=n$, in which case $I_n = (r_n,1]$.

CSM will determine a decision on $H$ as follows.  We take samples
until $\alpha \notin I_n$, leading to the stopping time 
$$\tau=\inf\{n\in \N: \alpha\notin I_n\}.$$
If $I_n\subseteq [0,\alpha]$ we reject $H$.  If
$I_n\subseteq (\alpha,1]$ we do not reject $H$.  By construction, the
uniform bound on the resampling risk in \eqref{eq:unifboundRR} holds true.

It is not necessary to compute the roots of $(n+1)b(n,p,x) = \epsilon$
explicitly to check the stopping criterion. Indeed, by the initial
definition of $I_n$,
$$\tau = \inf \left\{ n \in \N: (n+1) b(n,\alpha,S_n) \leq \epsilon \right\},$$
which is computationally easier to check. 

For comparisons with SIMCTEST, it will be useful to write $\tau$ equivalently as
$$\tau=\inf\{n\in \N: S_n\geq u_n \text{ or }S_n\leq l_n\},$$
where $u_n=\max\{k:(n+1)b(n,\alpha,k)>\epsilon\}+1$ and
$l_n=\min\{k:(n+1)b(n,\alpha,k)>\epsilon\}-1$ for any $n \in \N$.  We call
$(l_n)_{n\in \N}$ and $(u_n)_{n\in \N}$ the (implied) stopping
boundaries.  Figure~\ref{fig:dif_alpha_bounds} illustrates the implied
stopping boundaries for different testing thresholds $\alpha$. 

We may also be required to provide a confidence interval of the true p-value when the CSM stops. This confidence interval can be straightforwardly estimated based upon Monte Carlo or bootstrap samples \citep{ruxton2013improving}. However, in CSM, this is even simpler as a confidence interval can be directly obtained by using $I_{\tau}$.

In order to define the resampling risk for CSM we define an estimator $\hat p_{\text{c}}$ of $p$ as
\begin{align}
\label{CSM_estimator}
\hat{p}_{\text{c}} = \begin{cases}
\frac{S_{\tau}}{\tau} & \tau < \infty,\\
\alpha & \tau = \infty.
\end{cases}
\end{align}

The following theorem shows that the resampling risk of CSM is uniformly bounded.
\begin{theorem}
\label{theorem_CSM}
The estimator $\hat p_{\text{c}}$ satisfies
$$ \sup_{p \in [0,1]} \text{RR}_p(\hat{p}_{c}) < \epsilon.  $$
\end{theorem}
The proof of Theorem~\ref{theorem_CSM} can be found in Appendix~\ref{section_proof}.

Will CSM stop in a finite number of steps?  If $p=\alpha$ then the
algorithm will only stop with probability of at most
$\epsilon$. Indeed,
$\PP_\alpha(\tau<\infty)=\PP_\alpha(\exists n\in \N: \alpha\notin
I_n)\leq \epsilon$ by construction.
However, if $p\neq \alpha$ then the algorithm will stop in a finite
number of steps with probability one.
\citet{Lai1976} shows that with probability one,
$\lim_{n\to\infty}f_n(S_n)=p=\lim_{n\to\infty}g_n(S_n)$ given $p\neq \alpha$,
thus implying the existence of $n\in \N$ such that $\alpha\notin I_n$. 

\begin{figure}
\centering
\resizebox*{10cm}{!}{\includegraphics{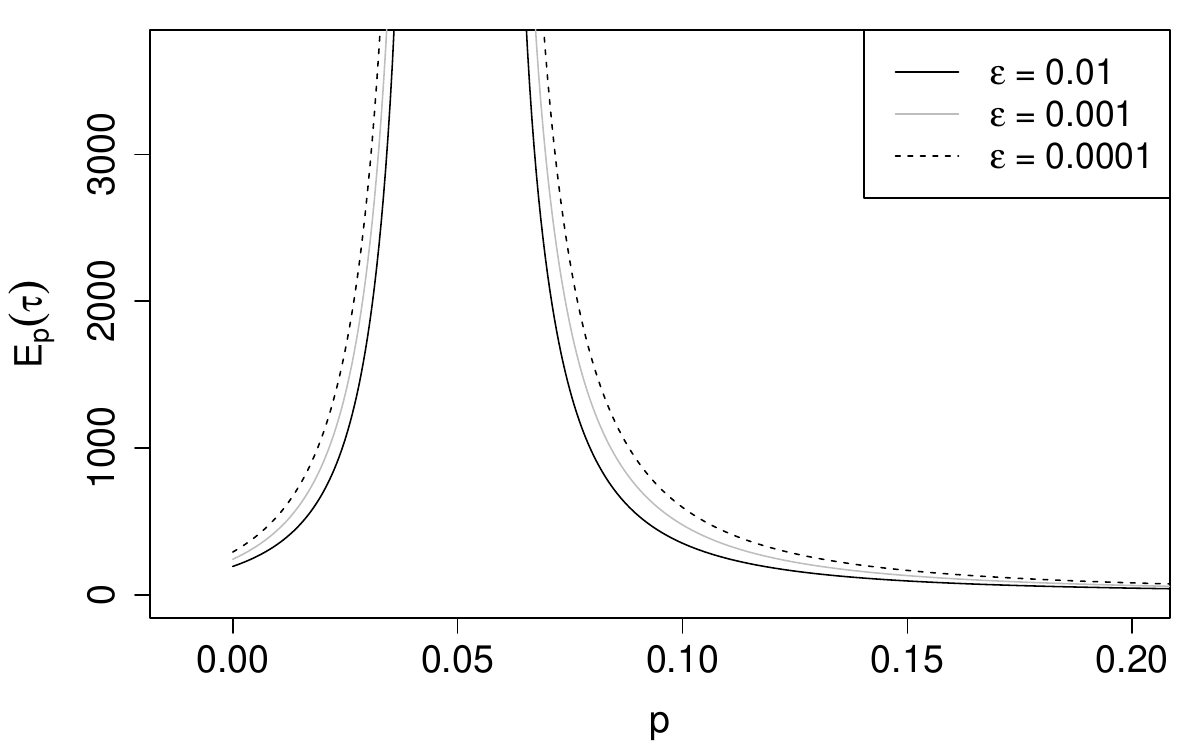}}\hspace{5pt}
\caption{Expected number of steps $\E_{p}(\tau)$ required to decide whether $p$ lies above or below the threshold $\alpha = 0.05$.} 
\label{fig:expeff}
\end{figure}

Figure~\ref{fig:expeff} shows the expected number of steps as a function of $p$
for three different values $\epsilon \in \{0.01, 0.001, 0.0001\}$.
The testing threshold chosen for Figure~\ref{fig:expeff} was $\alpha = 0.05$. In Figure~\ref{fig:expeff}, the expected number of steps increases slightly as the resampling risk decreases. For a fixed value of the resampling risk, the effort increases when the true p-value approaches the threshold $\alpha$.

If one considers a setup in which $p$ is random (e.g., for power/level
computations or in a Bayesian setup), then $\E[\tau]=\infty$.
This is a consequence of \cite[Equation 4.81]{wald1945} which also applies to any procedure
satisfying \eqref{eq:unifboundRR} for $\epsilon<0.5$
(see also \citep[Section 3.1]{Gandy2009}).
To have a feasible algorithm for a random $p$, a finite upper threshold on the
number of steps of CSM has to be imposed.  Alternatively, a
specialised procedure can be used (e.g., \cite{gandrub11}).

\section{Review of SIMCTEST}
\label{subsection_method_gandy}

\begin{figure}
\centering
\resizebox*{10cm}{!}{\includegraphics{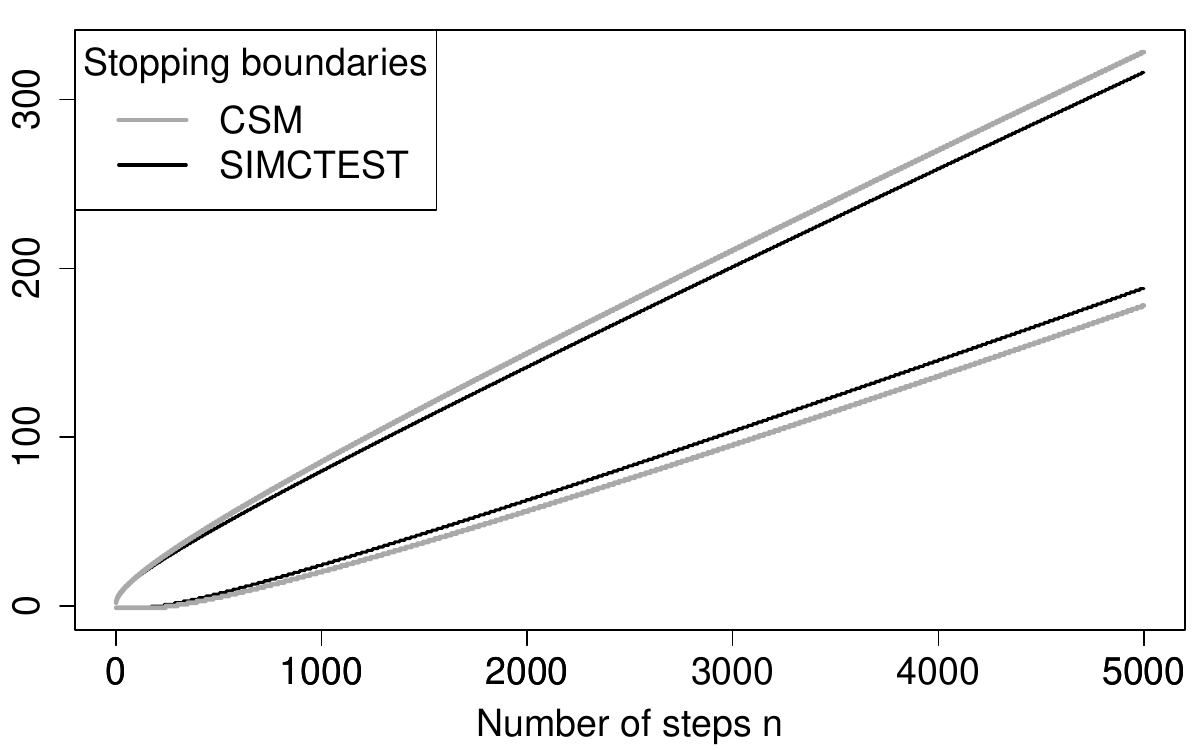}}\hspace{5pt}
\caption{Stopping boundaries for CSM and SIMCTEST (with default spending sequence).} 
\label{fig:5percent_bounds_graph}
\end{figure}

This section reviews the SIMCTEST method of \cite{Gandy2009}
(Sequential Implementation of Monte Carlo Tests)
which also bounds the resampling risk uniformly.
SIMCTEST sequentially updates
two integer sequences $(L_n)_{n \in \N}$ and
$(U_n)_{n \in \N}$ serving as lower and upper stopping boundaries and
stops the sampling process once the trajectory $(n,S_n)$ hits either boundary.
The decision whether the p-value lies above (below) the threshold
depends on whether the upper (lower) boundary was hit first.

The boundaries $(L_n)_{n \in \N}$ and $(U_n)_{n \in \N}$ are a
function of $\alpha$, computed recursively such that the probability of
hitting the upper (lower) boundary, given $p \leq \alpha$
($p > \alpha$), is less than $\epsilon$.  Starting with $U_1 = 2$,
$L_1 = -1$, the boundaries are recursively defined as
\begin{align*}
U_n = \min \{ j \in \N: &\PP_\alpha (\tau_\alpha \geq n, S_n \geq j) + \PP_\alpha (\tau_\alpha < n, S_\tau \geq U_\tau) \leq \epsilon_n \},\\
L_n = \max \{ j \in \Z: &\PP_\alpha (\tau_\alpha \geq n, S_n \leq j) + \PP_\alpha (\tau_\alpha < n, S_\tau \leq L_\tau) \leq \epsilon_n \},
\end{align*}
where $\epsilon_n$, $n \in \N$, is called a \textit{spending sequence}.  The
spending sequence is non-decreasing and satisfies
$\epsilon_n \rightarrow \epsilon$ as $n \rightarrow \infty$ as well as
$0 \leq \epsilon_n < \epsilon$ for all $n \in \N$.
Its purpose is to control how the
overall resampling risk $\epsilon$ is spent over all iterations of the algorithm:
In any step $n$ of SIMCTEST, new boundaries are computed with a risk of $\epsilon_n-\epsilon_{n-1}$.
\cite{Gandy2009} suggests $\epsilon_{n} = \epsilon n/(n + k)$ as a default spending sequence,
where $k$ is a constant,
and chooses $k = 1000$.

SIMCTEST stops as soon as the trajectory
$(n,S_n)$ hits the lower or upper boundary,
thus leading to the stopping time
$\sigma = \inf \{k \in \N: S_k \geq U_k \text{ or } S_k \leq L_k \}$.
In this case,
a p-value estimate can readily be computed as
$\hat{p}_{\text{s}}=S_{\sigma}/\sigma$ if $\sigma<\infty$ and $\hat{p}_{\text{s}}=\alpha$ otherwise.
Similarly to Figure~\ref{fig:expeff},
the expected stopping time of SIMCTEST diverges as $p$ approaches the threshold $\alpha$.

SIMCTEST achieves the desired uniform bound on the resampling risk
under certain conditions. To be precise, Theorem 2
in \cite{Gandy2009} states that if $\epsilon \leq 1/4$ and
$\log(\epsilon_n - \epsilon_{n-1}) = o(n)$ as $n \rightarrow \infty$,
then \eqref{eq:unifboundRR} holds with $\hat p=\hat p_s$.

\section{Comparison of CSM to SIMCTEST with the default spending sequence}
\label{section_stopping_boundaries}
In this section we compare the asymptotic behaviour of the width of the stopping boundaries for CSM and SIMCTEST.
SIMCTEST is employed with the default spending sequence given in Section~\ref{subsection_method_gandy}.
Unless stated otherwise,
we always consider the threshold $\alpha=0.05$ and aim to control the resampling risk at $\epsilon = 10^{-3}$.

\subsection{Comparison of boundaries}
\begin{figure}
\centering
\resizebox*{10cm}{!}{\includegraphics{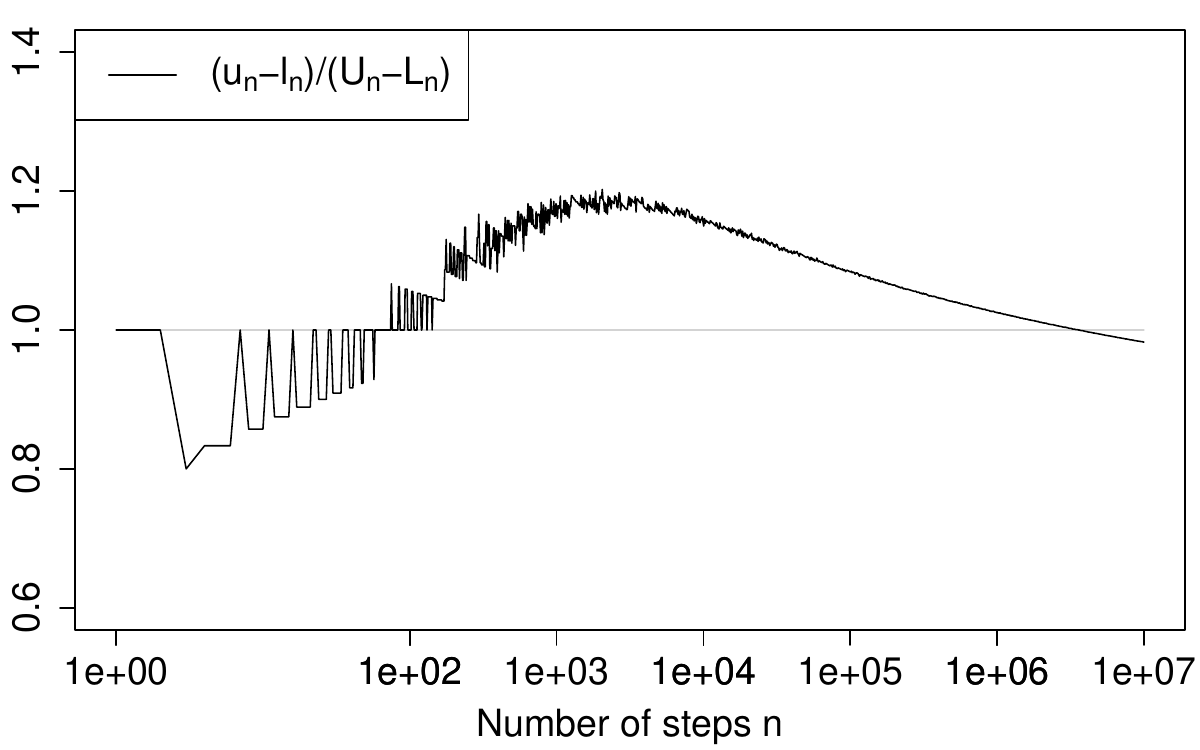}}\hspace{5pt}
\caption{Ratio of widths of stopping boundaries $(u_n-l_n)/(U_n-L_n)$ for CSM ($u_n$ upper, $l_n$ lower) and SIMCTEST with default spending sequence ($U_n$, $L_n$). Log scale on the x-axis.} 
\label{fig:ratio_bounds}
\end{figure}

We start by comparing the stopping boundaries of CSM and SIMCTEST.
First, Figure~\ref{fig:5percent_bounds_graph}
gives an overview of the upper and lower stopping boundaries of CSM and SIMCTEST up to 5000 steps, respectively.
Second, Figure~\ref{fig:ratio_bounds} shows the ratio of the widths of the
stopping boundaries for both methods, that is $(u_n-l_n)/(U_n-L_n)$, up to $10^{7}$ steps,
where $u_n$, $l_n$ ($U_n$, $L_n$) are the upper and lower stopping boundaries of CSM (SIMCTEST).

According to Figure~\ref{fig:ratio_bounds},
the boundaries of CSM are initially tighter than the ones of SIMCTEST
but become wider as the number of steps increases.
However, this will eventually reverse again for large numbers of steps as depicted in Figure~\ref{fig:ratio_bounds}.

\subsection{Real resampling risk in CSM and SIMCTEST} 
\label{TES} 
Both SIMCTEST and CSM are guaranteed to bound the resampling risk by some constant
$\epsilon$ chosen in advance by the user.
We will demonstrate in this section that the actual resampling risk
(that is the actual probability of hitting a boundary leading to a wrong decision in any run of an algorithm)
for SIMCTEST is close to $\epsilon$, whereas CSM does not
make full use of the allocated resampling risk.
This in turn indicates that it might be
possible to construct boundaries for SIMCTEST which are uniformly tighter than the
ones of CSM; we will pursue this in Section~\ref{section_dominating_seq}.

We compute the actual resampling risk recursively for both methods by
calculating the probability of hitting the upper or the lower stopping
boundary in any step for the case $p = \alpha$; other values of
$p$ give smaller resampling risks.  This can be done as follows:
Suppose we know the distribution of $S_{n-1}$
conditional on not stopping up to step $n-1$.  This allows us to
compute the probability of stopping at step $n$ as well as to work out
the distribution of $S_{n}$ conditional on not stopping up to step $n$.

\begin{figure}
\centering
\resizebox*{10cm}{!}{\includegraphics{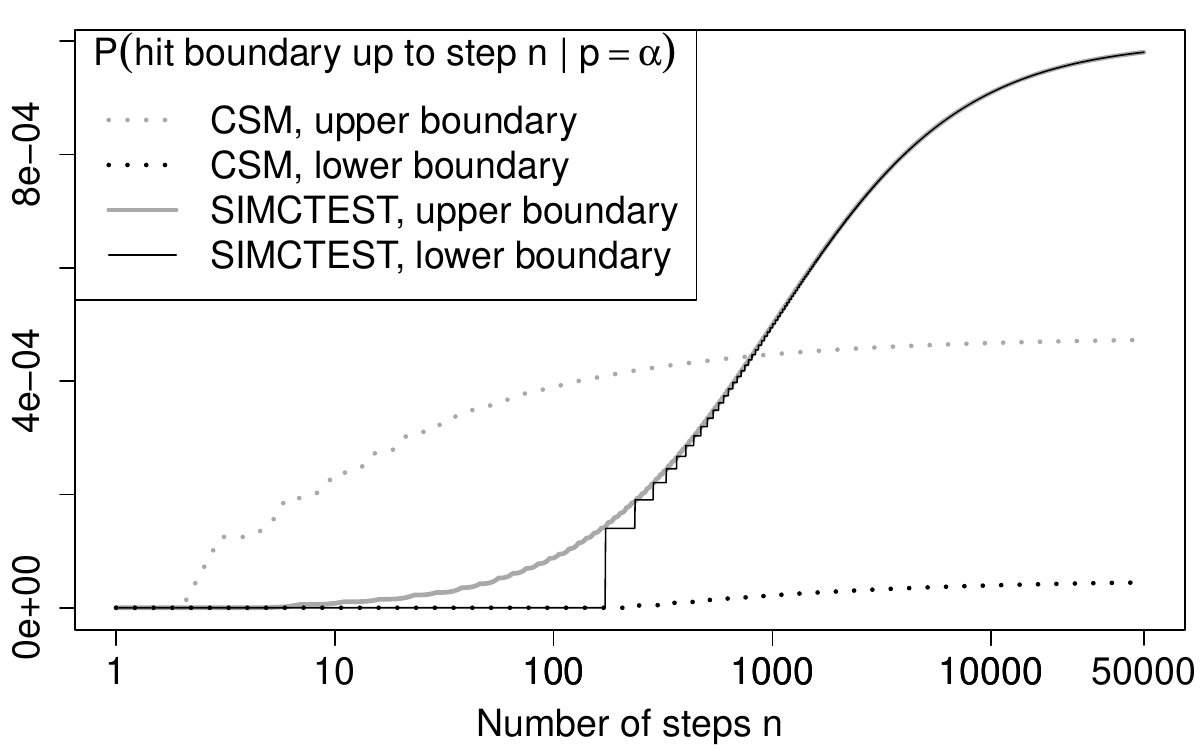}}\hspace{5pt}
\caption{Cumulative resampling risk spent over all iterations of the algorithm spent in CSM and SIMCTEST. Log scale on the x-axis.} 
\label{fig:hittingprobs}
\end{figure}

Figure~\ref{fig:hittingprobs} plots the cumulative probability of hitting the
upper and lower boundaries over $5\cdot 10^4$ steps for both methods.
As before we control the resampling risk at our default choice of $\epsilon = 10^{-3}$.

SIMCTEST seems to spend the full resampling risk as the number of
samples goes to infinity.  Indeed, the total probabilities of hitting
the upper and lower boundaries in SIMCTEST are both
$9.804 \cdot 10^{-4}$ within the first $5\cdot 10^4$ steps. This matches
the allowed resampling risk up to that point
of $\epsilon_{50000}=(5 \cdot 10^4)/(5 \cdot 10^4+1000) \epsilon \approx 9.804 \cdot 10^{-4}$
allocated by the spending sequence, which is close to the full resampling risk $\epsilon = 10^{-3}$.

CSM tends to be more
conservative as it does not spend the full resampling risk.  
Indeed, the total probabilities of hitting the upper and lower boundaries in CSM up to step $5 \cdot 10^4$
are $4.726 \cdot 10^{-4}$ and $4.472 \cdot 10^{-5}$, respectively. 
In particular, the probability of hitting the lower boundary in CSM is far less than $\epsilon$.  

This imbalance is more pronounced for even smaller thresholds.
We repeated the computation depicted in Figure~\ref{fig:hittingprobs}
for $\alpha \in \{0.02, 0.01, 0.005\}$ (figure not included),
confirming that the total probabilities of hitting the upper and lower boundaries in CSM
both decrease monotonically as $\alpha$ decreases.

\section{Spending sequences which dominate CSM}
\label{section_dominating_seq}

\subsection{Example of a bespoke spending sequence}
\label{Com_CSM_SIMCTEST_self}
One advantage of SIMCTEST lies in the fact that it allows control over the resampling
risk spent in each step through suitable adjustment of its spending sequence $\epsilon_{n}$, $n \in \N$.
This can be useful in practical
situations in which the overall computational effort is limited.  In
such cases, SIMCTEST can be tuned to spend the full resampling risk
over the maximum number of samples.  On the contrary, CSM has no
tuning parameters and hence does not offer a way to influence how the available
resampling risk is spent.

\begin{figure}
\centering
\resizebox*{10cm}{!}{\includegraphics{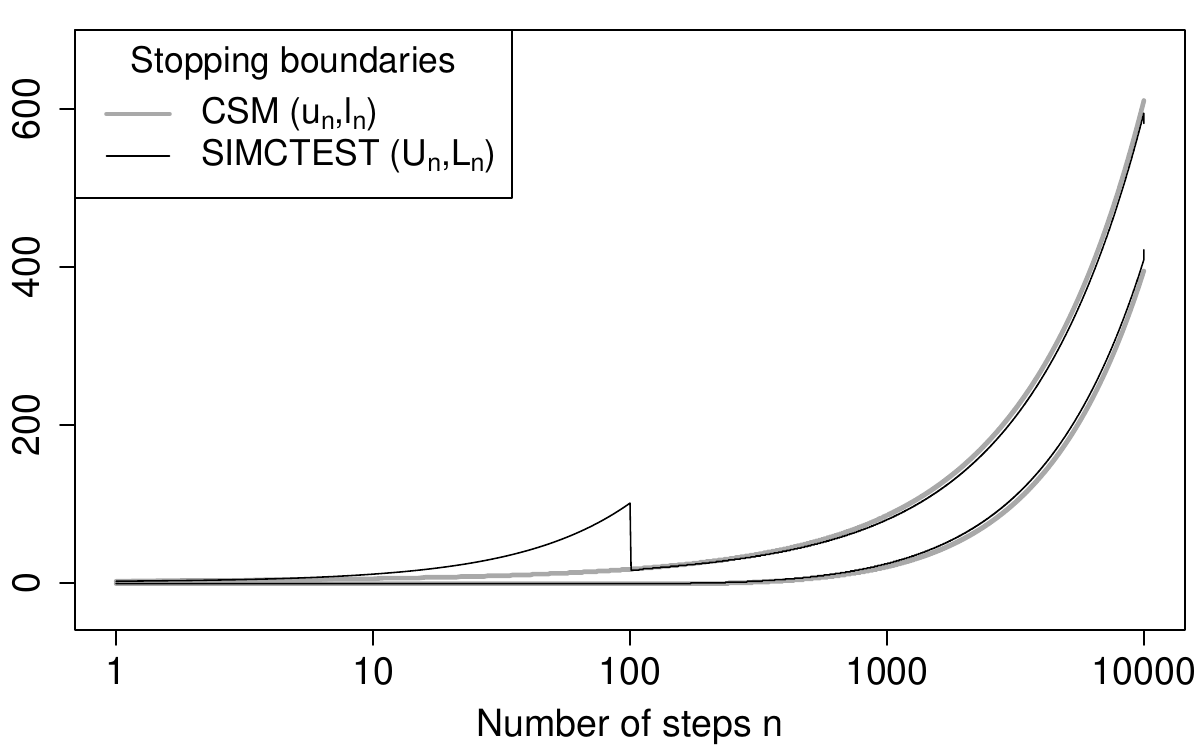}}\hspace{5pt}
\caption{Stopping boundaries of CSM and SIMCTEST with the truncated spending sequence. Log scale on the x-axis.} 
\label{5percent_bounds_new_spend_graph}
\end{figure}

Suppose we are given a lower bound $\mathcal{L}$
and an upper bound $\mathcal{U}$ for the minimal and maximal number of samples to be spent, respectively.
We construct a new spending sequence in SIMCTEST which guarantees
that no resampling risk is spent over both the first $\mathcal{L}$ samples
as well as after $\mathcal{U}$ samples have been generated.
We call this the \textit{truncated} spending sequence:
\begin{equation*}
  \epsilon_{n} =   
\begin{cases}
  0 &\text{ if } n \leq \mathcal{L},\\
  \frac{n}{n+k} \epsilon &\text{ if } \mathcal{L} < n <
  \mathcal{U},
  \\
  \epsilon&\text{ if } n \geq \mathcal{U}.
\end{cases}
\end{equation*}
Figure~\ref{5percent_bounds_new_spend_graph} shows
upper and lower stopping boundaries of CSM and SIMCTEST
with the truncated spending sequence (using $\mathcal{L} = 100$, $\mathcal{U} = 10000$, and $k=1000$).

As expected,
for the first $100$ steps the stopping boundaries of SIMCTEST are much wider than the ones of CSM since no resampling risk is spent.

As soon as SIMCTEST starts spending resampling risk on the computation of its stopping boundaries, the upper boundary drops considerably.
By construction,
the truncated spending sequence is chosen in such a way as to make SIMCTEST spend all resampling risk within $10^4$ steps.
Indeed, we observe in Figure~\ref{5percent_bounds_new_spend_graph} that as expected,
the stopping boundaries of SIMCTEST are uniformly narrower than those of CSM over the interval $(\mathcal{L},\mathcal{U})$,
thus resulting in a uniformly shorter stopping time for SIMCTEST.

We also observe, however, that this improvement in the width of the stopping boundaries seems to be rather marginal,
making the tuning-free CSM method a simple and appealing competitor.

\subsection{Uniformly dominating spending sequence}
\label{section_uniformly_dominating_seq}
Section~\ref{Com_CSM_SIMCTEST_self} gave an example in which it was
possible to choose the spending sequence for SIMCTEST in such a way as
to obtain stopping boundaries which are strictly contained within the
ones of CSM for a pre-specified range of steps.

Motivated by Figure~\ref{fig:hittingprobs} indicating that CSM does not spend the full resampling risk,
we aim to construct a spending sequence with the property
that the resulting boundaries in SIMCTEST are strictly contained within
the ones of CSM for every number of steps.
This implies that the stopping time of SIMCTEST is never longer than the one of CSM.
Our construction is dependant on the specific choice $\alpha = 0.05$.

We define a new spending sequence for SIMCTEST as $\epsilon_n = n^{0.5}/(n^{0.5} + 3) \cdot \epsilon$, $n \in \N$.
Appendix~\ref{Find_dom_seq} shows that this sequence empirically matches the rate at which SIMCTEST spends the real resampling risk with the one of CSM.

\begin{figure}
\centering
\resizebox*{10cm}{!}{\includegraphics{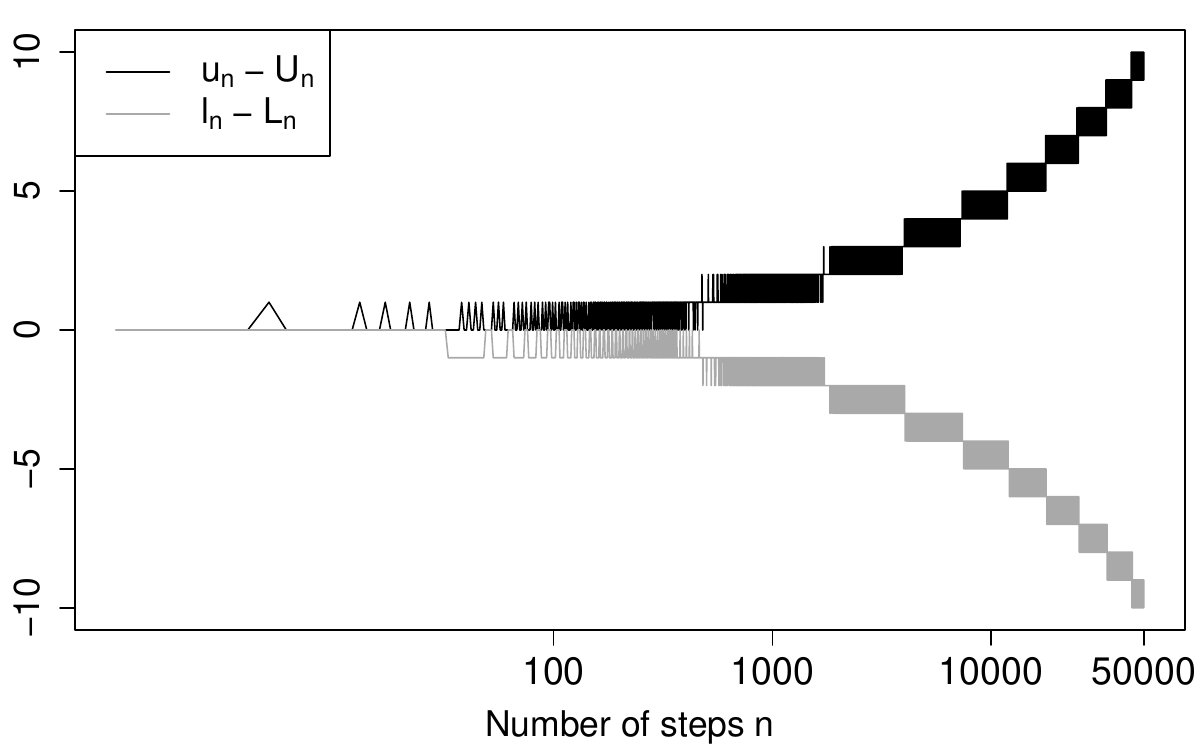}}\hspace{5pt}
\caption{Differences between the upper and lower stopping boundaries of CSM and SIMCTEST. Log scale on the x-axis.} 
\label{Diff_bounds_CSM_SIMTEST}
\end{figure}

Figure~\ref{Diff_bounds_CSM_SIMTEST} depicts the differences between the upper and lower boundaries
of CSM (upper $u_n$, lower $l_n$) and SIMCTEST (upper $U_n$, lower $L_n$) with the aforementioned spending sequence.
We observe that $l_n \leq L_n$ as well as $u_n \geq U_n$ for $n \in \{1, \ldots, 5 \cdot 10^4\}$,
thus demonstrating that SIMCTEST can be tuned empirically to spend the resampling risk at the same rate as CSM
while providing strictly tighter upper and lower stopping boundaries (over a finite number of steps).
We observe that the gap between the boundaries seems to increase with the number of steps, leading to the conjecture that SIMCTEST has tighter boundaries for all  $n \in \N$.

\section{Comparison with other truncated sequential Monte Carlo procedures}
\label{section_RR_comparison}
In this section, we compute the resampling risk for several truncated
procedures as a function of $p$ and thus demonstrate that they do not
bound the resampling risk uniformly. 

We consider truncated versions of CSM and SIMCTEST, which we denote by tCSM and tSIMCTEST, as well as the
algorithms in \cite{BesagClifford1991,davidson2000bootstrap,Fay2007,silva2018truncated}.  The maximum number of samples is set to $13000$ to
roughly fit in with the maximum number of steps of
\cite{davidson2000bootstrap}. The methods of \cite{BesagClifford1991,davidson2000bootstrap,Fay2007,silva2018truncated} have tuning parameters,
which we choose as follows. As suggested in \cite{BesagClifford1991},
the recommended parameter $h$ controlling the number of exceedances is
set to $10$. For \cite{davidson2000bootstrap}, the pretest level
$\beta$ concerning the level and power of the test is set to $0.05$,
and the minimum and maximum number of simulations are 99 and 12799 as
suggested by the authors. For \cite{Fay2007}, we choose one set of
tuning parameters
$(p_{\alpha}, p_{0}, \alpha_{0}, \beta_{0}) = (0.04, 0.0614, 0.05,
0.05)$ recommended by the authors. For the algorithm of \cite[Section
4]{silva2018truncated}, a grid search on the parameters
$(m, s, t_{1}, C_{e})$ with the aim to bound the type I error at
$0.05$ and the global power loss at 21\% produces the values
$(13000,2,16,702)$.  The resampling risk parameter $\epsilon$ in
SIMCTEST and CSM is set to $0.05$.

\begin{figure}
\centering
\resizebox*{10cm}{!}{\includegraphics{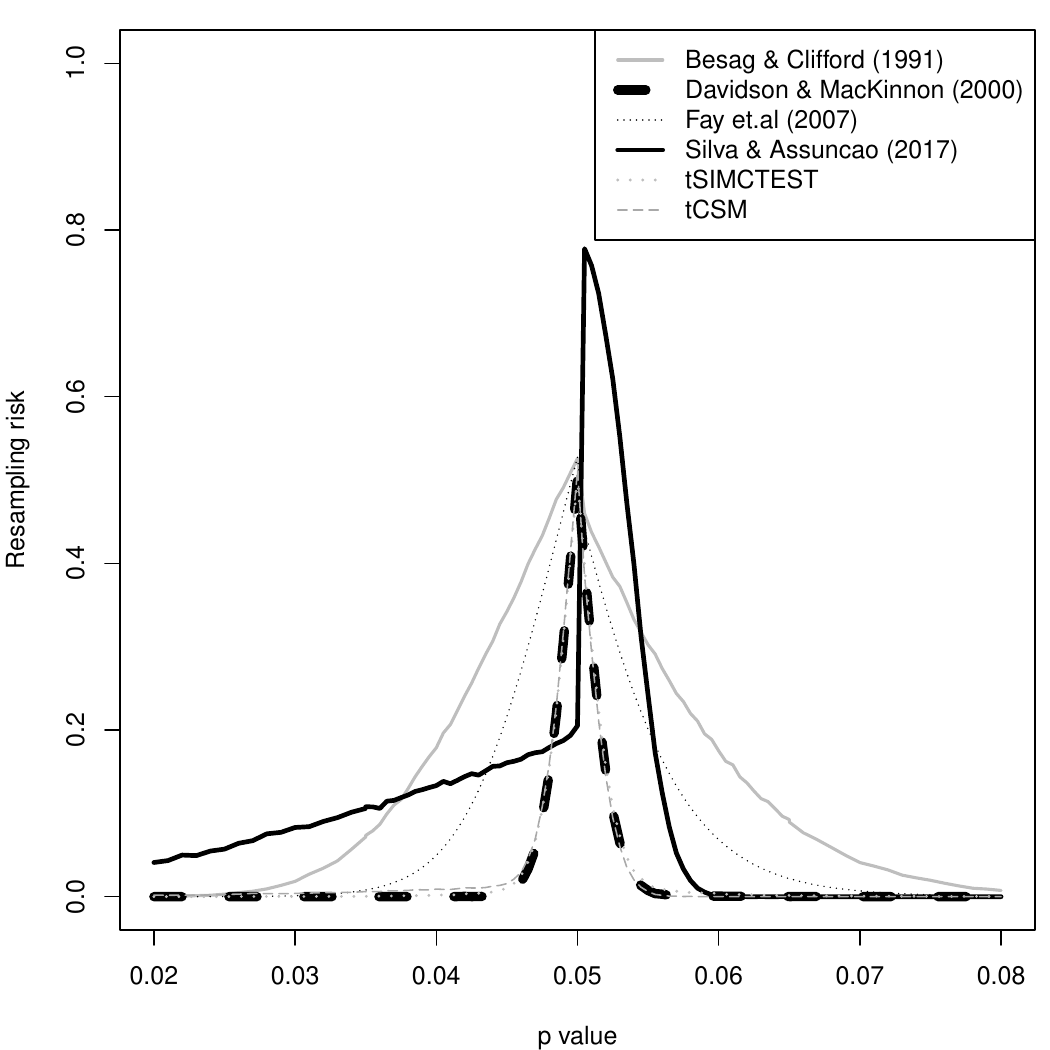}}\hspace{5pt}
\caption{Resampling risk incurred by several  truncated procedures as a function of the underlying $p$. Threshold $\alpha = 0.05$.}
\label{fig:RR_pvalue}
\end{figure}

Figure \ref{fig:RR_pvalue} shows the resampling risk as a function of the p-value. As expected, a truncated test results in a resampling risk of at least 50\% when $p = \alpha$. For other p-values than $p=\alpha$, the resampling risk can be smaller depending on the type of algorithm used and the number of samples it draws. However, tCSM is still amongst the best performers as it yields a low resampling risk almost everywhere with a localised spike at $0.5$. It also guarantees to bound the risk uniformly when the number of simulations tends to infinity while other truncated procedures \cite{BesagClifford1991, davidson2000bootstrap, Fay2007} do not have this property.

\section{Application}
\label{section_example}

\subsection{Comparison of penguin counts on two islands}
We apply CSM and SIMCTEST to a real data example which compares the number of breeding yellow-eyed penguin pairs on two types of islands: Stewart Island (on which cats are the natural predators of penguins) and some \textit{cat-free} islands nearby \citep{massaro2003comparison}. The number of yellow-eyed penguin pairs are recorded in $19$ discrete locations on Steward Island, resulting in an average count of $4.2$ and the following individual counts per location:
$$\{7,3,3,7,3,7,3,10,1,7,4,1,3,2,1,2,9,4,2\}.$$
Likewise, counts at $10$ discrete locations on the cat-free islands yield an average count of $9.9$ and individual counts of
$$\{15,32,1,13,14,11,1,3,2,7\}.$$

\cite{ruxton2013improving} employ SIMCTEST to conduct a hypothesis test to determine whether the means of the penguin counts on Stewart Island are equal to the ones of the cat-free islands. They apply Welch's $t$-test \citep{welch1947generalization} to assess whether two population groups have equal means. The test statistic of Welch's $t$-test is given as

$$T = \frac{\hat{\mu}_{1} - \hat{\mu}_{2}}{\sqrt{\frac{s_{1}^2}{n_{1}} + \frac{s_{2}^2}{n_{2}}}},$$
where $n_{1}, n_{2}$ are the sample sizes, $\hat{\mu}_{1}, \hat{\mu}_{2}$ are the sample means and $s_{1}^2$, $s_{2}^2$ are the sample variances of the two groups.

Under the assumption of normality of the two population groups, the distribution of the test statistic $T$ under the null hypothesis is approximately a Student's $t$-distribution with $v$ degrees of freedom, where

$$v = \frac{   \bigg(\frac{1}{n_{1}}  + \frac{s^{2}_{2}   }{   s_{1}^{2}  n_{2} }  \bigg)^{2} }{  \frac{1}{ n_{1}^{2} (n_{1} - 1) } + \frac{s_{2}^{2}}{ s_{1}^{2} n_{2}^{2} (n_{2} - 1)  }   }.$$
Using the above data, we obtain $t = -0.45$ as the observed test statistic and a p-value of 0.09.

As the normality assumption may not be satisfied in our case and as the $t$-distribution is only an approximation, \cite{ruxton2013improving} implement a parametric bootstrap test which randomly allocates each of the 178 penguin pairs to one of the 29 islands, where each island is chosen with equal probability. Based on their experiments, they conclude that they cannot reject the null hypothesis at a 5\% level.

Likewise, we apply CSM and SIMCTEST with the same bootstrap sampling procedure. We record the average effort measured in terms of the total number of samples generated. We set the resampling risk to $\epsilon = 0.001$ and use SIMCTEST with its default spending sequence $\epsilon_{n} = \epsilon n/(n+1000)$. 

We first perform a single run of both CSM and SIMCTEST. CSM and SIMCTEST stop after 751 and 724 steps with p-value estimates of 0.09 and 0.08, respectively. Hence, both algorithms reject the null hypothesis. We then conduct $10000$ independent runs to stabilise the results. Amongst those $10000$ runs, CSM rejects the null hypothesis $10000$ times compared with $9999$ times for SIMCTEST.  The average efforts of CSM and SIMCTEST are 1440 and 1131, respectively. Therefore, in this example, CSM gives comparable results to SIMCTEST while generating more samples on average.  We expect such behaviour due to the wider stopping boundaries of CSM in comparison with SIMCTEST (see Figure~\ref{fig:5percent_bounds_graph}).  However, we need to pre-compute the stopping boundaries of SIMCTEST in advance, which is not necessary in CSM.

\subsection{Autocorrelation in the sunspot time series}
We investigate the performance of CSM and SIMCTEST in a real time series for testing lag autocorrelation based upon the generalised Durbin--Watson test \citep{vinod1973generalization}. The test is designed to detect the autocorrelation at some lag $k$ in the residuals from regression analysis.  The null hypothesis asserts that the autocorrelation at lag $k$ is zero and the test statistic $d_{k}$ is defined by
$$d_{k} = \sum_{t = k+1}^{n} (y_{t} - y_{t-k})^2 / \sum_{t = 1}^{n} (y_{t} - \bar{y})^{2},$$  
where  we denote the time series by $\{y_{t}\}_{t = 1}^{n}$ and $\bar{y} = \frac{1}{n} \sum_{t = 1}^{n}$.
The null distribution of $d_{k}$ is usually analytically intractable except for some special cases, for example, when $y_{t}$ is normally distributed \citep{ali1984distributions}. Different methods are proposed for approximating the null distribution \citep{sneek1983some, ali1984distributions, ali1987durbin}. 

The time series we are interested in consists of the sunspot number per year from 1770 to 1869 \cite[Series E]{box2015time}. Applying the generalised Durbin--Watson test \citep{vinod1973generalization} with different approximation techniques of the null distribution, \cite{ali1984distributions, ali1987durbin} obtain different significance results regarding the lag autocorrelations. To be more precise, \cite{ali1984distributions} detects significant autocorrelations only at lag 1,2,9,10 at a 5\% level whereas \cite{ali1987durbin} finds that significant autocorrelations at lag 5,6,11 and 12 also exist. 

To employ CSM and SIMCTEST for testing the lag autocorrelations, we use the approach proposed by \cite{mackinnon2002bootstrap} to generate bootstrap samples. We first calculate the residuals $\epsilon_{t} = y_{t} - \bar{y}$ for $t = 1, \ldots, n$. To simulate a new bootstrap sample, we let $y^{*}_{t} = \bar{y} + \epsilon^{*}_{t}$ for $t = 1, \ldots, n$, where each $\epsilon^{*}_{t}$ is randomly chosen from $\{\epsilon_{t}\}_{t = 1}^{n}$ with equal probability. Given the bootstrap sample $\{y^{*}_{t}\}_{t=1}^{n}$, we can compute the corresponding test statistic $d_{k}$ and compare it to the observed test statistic. 

We run CSM and SIMCTEST following the aforementioned bootstrap sampling procedure. We set the resampling risk $\epsilon = 0.001$ in both algorithms, and the spending sequence $\epsilon_{n} = \epsilon n/(n+1000)$ in SIMCTEST. After a single run of CSM and SIMCTEST (based on the same bootstrap samples), we obtain significant results at a 5\% level at lag 1,2,5,6,9,10,11,12, which coincide with \cite{ali1987durbin}. The stopping time for CSM and SIMCTEST is shown in Table~\ref{sunspot}. An early stopping time of CSM with fewer than 30 bootstrap samples generated usually implies a slightly faster algorithm than SIMCTEST, as can be seen at lag 7,14 and 15. These lags do not reject the null hypothesis. For other lags, in particular those which reject the null hypothesis, SIMCTEST enjoys an earlier stopping time than CSM.
\begin{table}[t]
\label{sunspot}
\centering
\resizebox{\textwidth}{!}{
\begin{tabular}{|c|r|r|r|r|r|r|r|r|r|r|r|r|r|r|r|r|r|}
\hline
Lag & 1 & 2 & 3 & 4 & 5 & 6 & 7 & 8 & 9 & 10 & 11 & 12 & 13 & 14 & 15\\\hline
CSM  & 520  & 520  & 78  & 1761  &  639   & 2375   &  25  & 170                                         & 520 & 520 & 520 & 520 & 1152 & 24 & 13\\\hline
SIMCTEST  & 329 & 329 & 76  & 772 & 329  & 1686 & 27                                                              & 164 & 329 & 329 & 329 & 329  & 981 & 25 & 20\\\hline
\end{tabular}
}
\caption{Stopping time of CSM and SIMCTEST for testing the lag autocorrelations of the residuals in the sunspot time series \cite[Series E]{box2015time}.}
\end{table}

\section{Discussion}
\label{section_discussion}

\begin{table}[t]
\centering
\label{comparison}
\resizebox{\textwidth}{!}{
\begin{tabular}{|l|l|l|p{4cm}|}
\hline
& CSM & SIMCTEST & SIMCTEST with pre-\\
	    &&&computed boundaries\\
\hline
Memory requirement                  & $O(1)$         & $O(\sqrt{\tau \log \tau})^{*}$    & $O(\tau_{\max})$                                                       \\\hline
Computational effort               & $O(\tau)$     & $O(\tau \sqrt{\tau \log \tau})^{*}$& $O(\tau)$                                                                       \\
\hline
Parameters of each method 
  & $\epsilon$            & $ \{ \epsilon_{n}  \}_{n \in \mathbb{N}}$                            & $ \{ \epsilon_{n}  \}_{n \in \mathbb{N}}$                                                                             \\
 \hline
Implementation from scratch        & Very easy& Easy                           & Easy\\
 \hline
\end{tabular}
}
\caption{Comparison between CSM and SIMCTEST.
The parameter $\tau$ denotes the stopping time and $\tau_{\max}$ denotes the maximum length of the pre-computed boundaries for SIMCTEST. Empirical quantities are denoted with $^{*}$.}
\end{table}

This article introduces a new method called CSM to decide whether an
unknown p-value, which can only be approximated via Monte Carlo
sampling, lies above or below a fixed threshold $\alpha$ while
uniformly bounding the resampling risk at a user-specified
$\epsilon>0$.  The method is straightforward to implement and relies
on the construction of a confidence sequence
\citep{Robbins1970,Lai1976} for the unknown p-value.

We compare CSM to SIMCTEST \citep{Gandy2009},
finding that CSM is the more conservative method:
The (implied) stopping boundaries of CSM are generally wider than the ones of SIMCTEST
and in contrast to SIMCTEST,
CSM does not fully spend the allocated resampling risk $\epsilon$.

We use these findings in two ways: First, an upper bound is usually
known for the maximal number of samples which can be spent in
practical applications.  We construct a \textit{truncated} spending
sequence for SIMCTEST which spends all the available resampling risk
within a pre-specified interval, thus leading to uniformly tighter
stopping boundaries and shorter stopping times than CSM.  Second, we
empirically analyse at which rate CSM spends the resampling risk.  By
matching this rate with a suitably chosen spending sequence, we
empirically tune the stopping boundaries of SIMCTEST to uniformly
dominate those of CSM even for open-ended sampling.

A comparison of memory requirement and computational effort for CSM
and SIMCTEST is given in Table~\ref{comparison}.  In SIMCTEST, the
boundaries are sequentially calculated as further samples are being
generated whereas in SIMCTEST with pre-computed boundaries, the
boundaries are initially computed and stored up to a maximum number of
steps $\tau_{\max}$.  In CSM, solely the cumulative sum $S_{n}$ needs
to be stored in each step, leading to a memory requirement of $O(1)$.
\citet{Gandy2009} empirically shows that SIMCTEST with the default
spending sequence has a memory requirement of
$O(\sqrt{\tau \log \tau})$.  In SIMCTEST with pre-computed boundaries and default spending sequence, 
the amount of memory required temporarily up to step $n$ is
$O(\sqrt{n \log n})$. To compute the boundaries up to $\tau_{\max}$, a
total memory of $O(\sqrt{\tau_{\max} \log \tau_{\max}})$ is hence
required. Additionally, the values of the upper and lower boundaries
up to $\tau_{\max}$ need to be stored, which requires $O(\tau_{\max})$
memory. Hence, the total memory requirement of SIMCTEST with
pre-computed boundaries is $O(\tau_{\max})$.  Evaluating the stopping
criterion in each step of CSM or SIMCTEST with pre-computed boundaries
requires $O(1)$, leading to the total computational effort of
$O(\tau)$ depicted in Table~\ref{comparison} for both cases.
The computational effort of SIMCTEST is roughly proportional to
$\sum_{n = 1}^{\tau} |U_{n} - L_{n}|$ \citep{Gandy2009}. Using the empirical result
$|U_{n} - L_{n}| \sim O(\sqrt{n \log n})$, we obtain a
bound of $O(\tau \sqrt{\tau \log \tau})$ for the computational effort
of SIMCTEST.

We also compare the truncated versions of CSM (tCSM) and SIMCTEST (tSIMCTEST) with other truncated sequential Monte Carlo procedures. We demonstrate empirically that the resampling risk of the truncated methods cannot be bounded by an arbitrary small number and exceeds 0.5 when the true p-value equals the threshold. Nevertheless, tCSM and tSIMCTEST are still among the best performers, meaning that they yield a low resampling risk almost everywhere with a localised spike at $0.5$.

The advantage of SIMCTEST (with pre-computed boundaries) lies in its adjustable spending sequence $\{\epsilon_{n}\}_{n \in \N}$: This flexibility allows the user to control the resampling risk spent in each step, thus enabling the user to spend no risk before a pre-specified step or to spend the full risk within a finite number of steps (see Section~\ref{Com_CSM_SIMCTEST_self}). This leads to (marginally) tighter stopping boundaries and faster decisions.
The strength of CSM, however, lies in its straightforward implementation compared to SIMCTEST. Both methods illustrate a superior performance (measured in resampling risk) if truncations are applied. Overall we conclude that the simplicity of CSM, and its comparable performance to SIMCTEST and other truncated procedures make it a very appealing competitor for practical applications.

\appendix
\section{Proof of Theorem~\ref{theorem_CSM}}
\label{section_proof}

\begin{proof}
We start by considering the case $p \leq \alpha$.
If $p\leq \alpha$, the resampling risk is
$\text{RR}_{p}(\hat{p}_{c}) = \PP_{p} (\hat{p}_{c} > \alpha)$.
We  show that $\hat{p}_{c} >\alpha$ only when hitting the
upper boundary and that the probability of hitting the upper
boundary is bounded by $\epsilon$.

To see the former: When not hitting any boundary, i.e.\ on the event $\{\tau =\infty\}$,
we have $\hat{p}_{c}= \alpha$.  When hitting the lower boundary,
i.e.\ on the event $\{\tau<\infty, S_{\tau}\leq l_{\tau}\}$, we have
$\hat{p}_{c}=S_{\tau}/\tau\leq l_{\tau}/\tau$.  It thus suffices
to show $l_n/n\leq \alpha$ for all $n\in \N$.

Let $n\in \N$. By \eqref{eqn:lai_original},
$\PP_{\alpha}(b(n,\alpha,S_n) > \frac{\epsilon}{n+1}) \geq 1- \epsilon$.
Hence, there exists $k$ such that $(n+1)b(n,\alpha,k)>\epsilon$.
Furthermore, $b(n, \alpha,x)$ has a maximum at $x = \lceil\alpha n\rceil$
or at $x=\lfloor{\alpha n}\rfloor$.  Thus, there exists a
$k\in \{ \lceil{\alpha n}\rceil, \lfloor{\alpha n}\rfloor\}$ such that
$(n+1)b(n,\alpha,k)>\epsilon$.  Hence, by the definition of $l_n$ we
have $l_n\leq \lceil{\alpha n}\rceil-1< \alpha n$.

To finish the proof for this case we show that the probability of hitting the upper boundary is
bounded by $\epsilon$, which can be done using \cite[Lemma 3]{Gandy2009} and \eqref{eqn:lai_original}:
\begin{align*}
\PP_{p}( \tau < \infty, S_{\tau} \geq  u_{\tau})
&\leq   \PP_{\alpha}( \tau < \infty, S_{\tau} \geq  u_{\tau})  \leq \PP_{\alpha}( \tau < \infty)\\ 
&=\PP_{\alpha}(\exists n\in \N: (n+1) b(n,\alpha,S_n) \leq \epsilon)\leq \epsilon.
\end{align*}

The case $p > \alpha$ can be proven analogously to the case $p<\alpha$
using that $\PP_p(\tau=\infty)=0$, which is shown in
\cite[p.\ 268]{Lai1976}.
\end{proof}

\section{Finding a uniformly dominating spending sequence}
\label{Find_dom_seq}
\begin{figure}
\centering
\resizebox*{7cm}{!}{\includegraphics{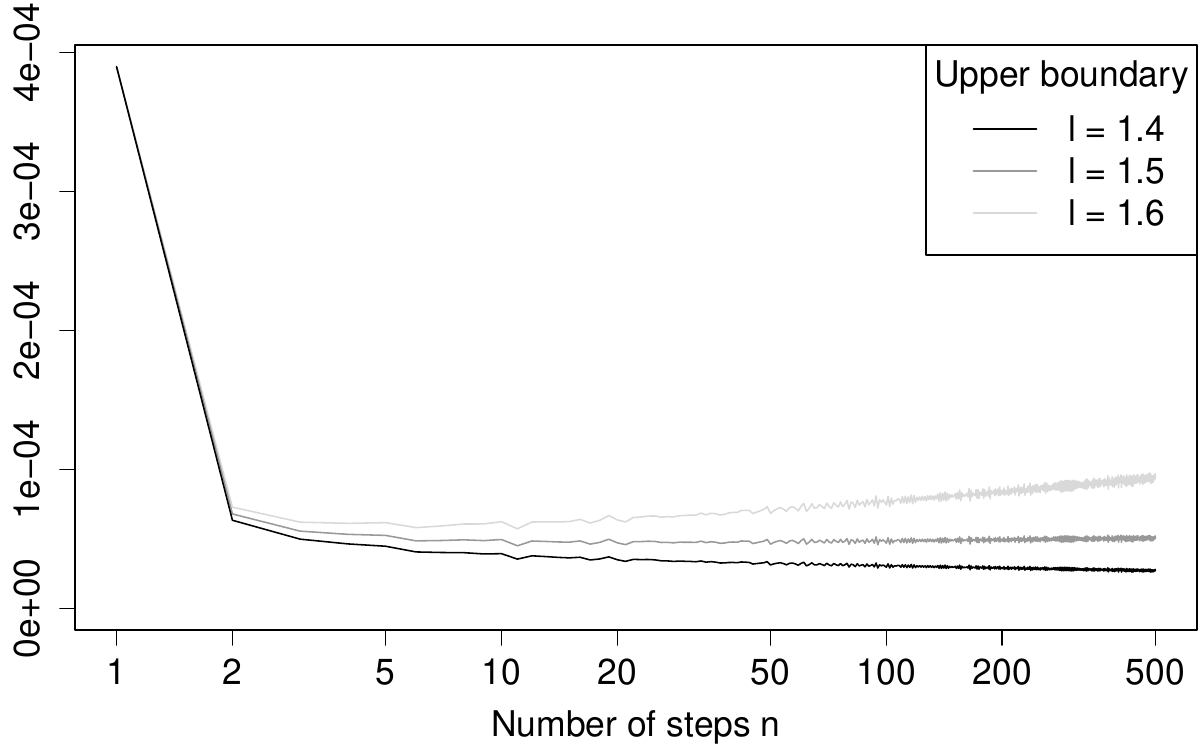}}\hspace{5pt}
\resizebox*{7cm}{!}{\includegraphics{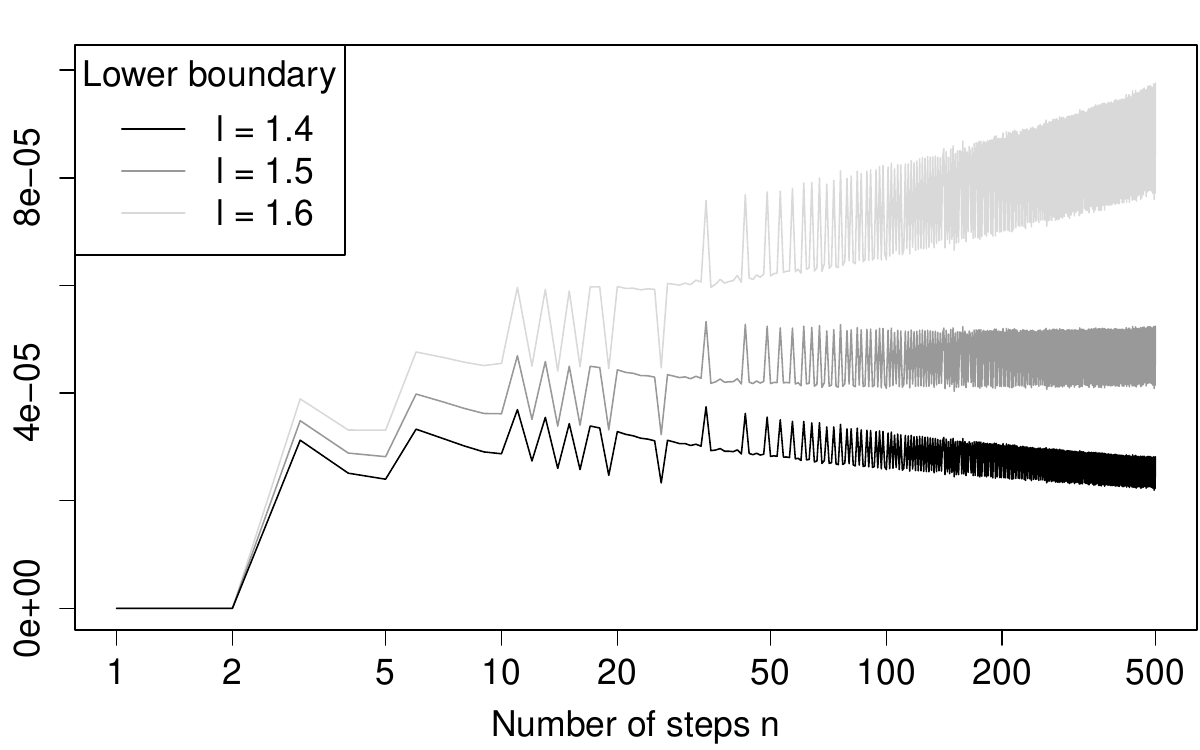}}
\caption{Trajectories of $\Delta \tilde{\epsilon}_{100n} n^l$
in the upper (left) and lower (right) boundary for $l = \{1.4, 1.5, 1.6\}$.
Log scale on the x-axis.} \label{Rate_cov_CSM}
\end{figure}

In Section~\ref{section_uniformly_dominating_seq}
we aim to find a spending sequence in SIMCTEST so that its boundaries are strictly contained within
the ones of CSM for every number of steps.

We achieve this by first determining the (empirical) rate at which the real resampling risk is spent in each step in CSM.
By matching this rate using a suitably chosen spending sequence,
we obtain upper and lower stopping boundaries for SIMCTEST which are uniformly narrower than the ones of CSM
(verified for the first $5\cdot 10^4$ steps).

We start by estimating the rate at which the real resampling risk is spent in CSM.
We are interested in empirically finding an $l \in \R$ such that
$n^l \cdot \left( \epsilon_n^\text{CSM} - \epsilon_{n-1}^\text{CSM} \right)$
is constant, where
$\epsilon_n^\text{CSM}$
is the (cumulative) real resampling risk
(the total probability of hitting either boundary) for the first $n$ steps in CSM.

Figure~\ref{Rate_cov_CSM} depicts
$n^l \cdot \left( \epsilon_n^\text{CSM} - \epsilon_{n-1}^\text{CSM} \right)$
for both the upper (left plot) and lower (right plot)
boundary of CSM as a function of the number of steps $n$ and for three values $l \in \{1.4,1.5,1.6\}$.
Based on Figure~\ref{Rate_cov_CSM} we estimate that CSM spends the resampling risk at roughly $O(n^{-1.5})$
for both the upper and lower boundaries.

We start with an analytical calculation of the rate at which SIMCTEST spends the resampling risk (as opposed to also estimating it).
The default spending sequence in SIMCTEST is $\epsilon_{n}^\text{S} = n/(n + k)\epsilon$, $n \in \N$.
Hence the resampling risk spent in step $n \in \N$ is
$\Delta \epsilon_{n}^\text{S} = \epsilon_n^\text{S} - \epsilon_{n-1}^\text{S} = k/((n+k)(n+k-1)) \sim n^{-2}$.
We conducted simulations (similar to the ones in Figure~\ref{Rate_cov_CSM} for CSM)
which indeed confirm the analytical $O(n^{-2})$ rate for SIMCTEST (simulations not included in this article).
Overall, the spending rate of $O(n^{-1.5})$ for CSM is thus slower than
the $O(n^{-2})$ rate of SIMCTEST with the default spending sequence.

In order to match the $O(n^{-1.5})$ rate for CSM,
we generalise the default spending sequence of SIMCTEST to
$\epsilon_n^\text{S} = n^\gamma/ \left( n^\gamma + k \right) \epsilon$ for $n \in \N$ and a fixed $\gamma>0$.
Similarly to the aforementioned derivation,
SIMCTEST in connection with $\epsilon_n^\text{S}$
will spend the real resampling risk at a rate of $O \left( n^{-(\gamma + 1)} \right)$.
We choose the parameters $\gamma$ and $k$ to obtain stopping boundaries for SIMCTEST which dominate the ones of CSM.
First, we set $\gamma=0.5$ to match the rate of CSM.
Second, we empirically determine $k$ to keep the stopping boundaries of SIMCTEST within the ones of CSM
(for the range of steps $n \in \{1,\ldots,5\cdot 10^4\}$ considered in Figure~\ref{Diff_bounds_CSM_SIMTEST}).
We find that the choice $k = 3$ satisfies this condition.


\begin{thebibliography}{}
\bibitem[Ali, 1984]{ali1984distributions}
Ali, M.~M. (1984).
\newblock Distributions of the sample autocorrelations when observations are
  from a stationary autoregressive-moving-average process.
\newblock {\em Journal of Business \& Economic Statistics}, 2(3):271--278.

\bibitem[Ali, 1987]{ali1987durbin}
Ali, M.~M. (1987).
\newblock Durbin--{W}atson and generalized {D}urbin--{W}atson tests for
  autocorrelations and randomness.
\newblock {\em Journal of Business \& Economic Statistics}, 5(2):195--203.

\bibitem[Besag and Clifford, 1991]{BesagClifford1991}
Besag, J. and Clifford, P. (1991).
\newblock Sequential {M}onte {C}arlo {$p$}-values.
\newblock {\em Biometrika}, 78(2):301--304.

\bibitem[Bonferroni, 1936]{Bonferroni1936}
Bonferroni, C. (1936).
\newblock Teoria statistica delle classi e calcolo delle probabilit\`a.
\newblock {\em {P}ubblicazioni del {R} {I}stituto {S}uperiore di {S}cienze
  {E}conomiche e {C}ommerciali di {F}irenze}, 8:3--62.

\bibitem[Box et~al., 2015]{box2015time}
Box, G.~E., Jenkins, G.~M., Reinsel, G.~C., and Ljung, G.~M. (2015).
\newblock {\em Time series analysis: forecasting and control}.
\newblock John Wiley \& Sons.

\bibitem[Davidson and MacKinnon, 2000]{davidson2000bootstrap}
Davidson, R. and MacKinnon, J.~G. (2000).
\newblock Bootstrap tests: How many bootstraps?
\newblock {\em Econometric Reviews}, 19(1):55--68.

\bibitem[Davison and Hinkley, 1997]{davison1997bootstrap}
Davison, A.~C. and Hinkley, D.~V. (1997).
\newblock {\em Bootstrap methods and their application}, volume~1 of {\em
  Cambridge Series in Statistical and Probabilistic Mathematics}.
\newblock Cambridge University Press.

\bibitem[Fay and Follmann, 2002]{FayFollmann2002}
Fay, M.~P. and Follmann, D.~A. (2002).
\newblock Designing {M}onte {C}arlo implementations of permutation or bootstrap
  hypothesis tests.
\newblock {\em The American Statistician}, 56(1):63--70.

\bibitem[Fay et~al., 2007]{Fay2007}
Fay, M.~P., Kim, H.-J., and Hachey, M. (2007).
\newblock On using truncated sequential probability ratio test boundaries for
  {M}onte {C}arlo implementation of hypothesis tests.
\newblock {\em Journal of Computational and Graphical Statistics},
  16(4):946--967.

\bibitem[Gandy, 2009]{Gandy2009}
Gandy, A. (2009).
\newblock Sequential implementation of {M}onte {C}arlo tests with uniformly
  bounded resampling risk.
\newblock {\em Journal of the American Statistical Association},
  104(488):1504--1511.

\bibitem[Gandy and Hahn, 2014]{GandyHahn2014}
Gandy, A. and Hahn, G. (2014).
\newblock M{MCT}est---a safe algorithm for implementing multiple {M}onte
  {C}arlo tests.
\newblock {\em Scandinavian Journal of Statistics. Theory and Applications},
  41(4):1083--1101.

\bibitem[Gandy and Hahn, 2016]{GandyHahn2016framework}
Gandy, A. and Hahn, G. (2016).
\newblock A framework for {M}onte {C}arlo based multiple testing.
\newblock {\em Scandinavian Journal of Statistics, Theory and Applications},
  43(4):1046--1063.

\bibitem[Gandy and Rubin-Delanchy, 2013]{gandrub11}
Gandy, A. and Rubin-Delanchy, P. (2013).
\newblock An algorithm to compute the power of {M}onte {C}arlo tests with
  guaranteed precision.
\newblock {\em The Annals of Statistics}, 41(1):125--142.

\bibitem[Gleser, 1996]{gleser1996bootstrap}
Gleser, L.~J. (1996).
\newblock Comment on "{B}ootstrap {C}onfidence {I}ntervals" by {D}i{C}iccio and
  {E}fron.
\newblock {\em Statistical Science}, 11(3):219--221.

\bibitem[Kim, 2010]{Kim2010}
Kim, H.-J. (2010).
\newblock Bounding the resampling risk for sequential {M}onte {C}arlo
  implementation of hypothesis tests.
\newblock {\em Journal of Statistical Planning and Inference},
  140(7):1834--1843.

\bibitem[Kulldorff, 2001]{kulldorff2001prospective}
Kulldorff, M. (2001).
\newblock Prospective time periodic geographical disease surveillance using a
  scan statistic.
\newblock {\em Journal of the Royal Statistical Society: Series A (Statistics
  in Society)}, 164(1):61--72.

\bibitem[Lai, 1976]{Lai1976}
Lai, T.~L. (1976).
\newblock On confidence sequences.
\newblock {\em The Annals of Statistics}, 4(2):265--280.

\bibitem[MacKinnon, 2002]{mackinnon2002bootstrap}
MacKinnon, J.~G. (2002).
\newblock Bootstrap inference in econometrics.
\newblock {\em Canadian Journal of Economics/Revue canadienne
  d'{\'e}conomique}, 35(4):615--645.

\bibitem[Massaro and Blair, 2003]{massaro2003comparison}
Massaro, M. and Blair, D. (2003).
\newblock Comparison of population numbers of yellow-eyed penguins,
  {M}egadyptes antipodes, on {S}tewart island and on adjacent cat-free islands.
\newblock {\em New Zealand Journal of Ecology}, pages 107--113.

\bibitem[Robbins, 1970]{Robbins1970}
Robbins, H. (1970).
\newblock Statistical methods related to the law of the iterated logarithm.
\newblock {\em Annals of Mathematical Statistics}, 41:1397--1409.

\bibitem[Ruxton and Neuh{\"a}user, 2013]{ruxton2013improving}
Ruxton, G.~D. and Neuh{\"a}user, M. (2013).
\newblock Improving the reporting of p-values generated by randomization
  methods.
\newblock {\em Methods in Ecology and Evolution}, 4(11):1033--1036.

\bibitem[Silva and Assun\c{c}\~ao, 2013]{SilvaAssuncao2013}
Silva, I. and Assun\c{c}\~ao, R. (2013).
\newblock Optimal generalized truncated sequential {M}onte {C}arlo test.
\newblock {\em Journal of Multivariate Analysis}, 121:33--49.

\bibitem[Silva and Assun\c{c}\~ao, 2018]{silva2018truncated}
Silva, I. and Assun\c{c}\~ao, R. (2018).
\newblock Truncated sequential {M}onte {C}arlo test with exact power.
\newblock {\em Brazilian Journal of Probability and Statistics},
  32(2):215--238.

\bibitem[Silva et~al., 2009]{Silva2009}
Silva, I., Assun\c{c}\~ao, R., and Costa, M. (2009).
\newblock Power of the sequential {M}onte {C}arlo test.
\newblock {\em Sequential Analysis. Design Methods \& Applications},
  28(2):163--174.

\bibitem[Sneek, 1983]{sneek1983some}
Sneek, J.~M. (1983).
\newblock {\em Some approximations to the exact distribution of sample
  autocorrelations for autoregressive moving average models}.
\newblock Time Series Analysis: Theory and Practice 3.

\bibitem[Tango and Takahashi, 2005]{tango2005flexibly}
Tango, T. and Takahashi, K. (2005).
\newblock A flexibly shaped spatial scan statistic for detecting clusters.
\newblock {\em International Journal of Health Geographics}, 4(1):11.

\bibitem[Vinod, 1973]{vinod1973generalization}
Vinod, H. (1973).
\newblock Generalization of the durbin-watson statistic for higher order
  autoregressive processes.
\newblock {\em Communications in Statistics-Theory and Methods}, 2(2):115--144.

\bibitem[Wald, 1945]{wald1945}
Wald, A. (1945).
\newblock Sequential tests of statistical hypotheses.
\newblock {\em Annals of Mathematical Statistics}, 16:117--186.

\bibitem[Welch, 1947]{welch1947generalization}
Welch, B.~L. (1947).
\newblock The generalization of `{S}tudent's' problem when several different
  population variances are involved.
\newblock {\em Biometrika}, 34:28--35.
\end{thebibliography}

\end{document}